\documentclass[manuscript]{acmart}

\AtBeginDocument{%
  }

\setcopyright{acmlicensed}
\copyrightyear{2025}
\acmYear{2025}
\acmDOI{XXXXXXX.XXXXXXX}
\acmConference[Working Paper in Progress]{Working Paper in Progress}{2024}{Online}
\acmISBN{978-1-4503-XXXX-X/2018/06}

\begin{document}

\title{(Working Paper) Good Faith Design: Religion as a Resource for Technologists}


\author{Nina Lutz}
\email{ninalutz@uw.edu}
\authornote{Work conducted during Microsoft Research Internship.}
\affiliation{
 \institution{University of Washington}
 \city{Seattle}
 \state{Washington}
 \country{USA}
}
\affiliation{
 \institution{Microsoft Research}
 \city{Redmond}
 \state{Washington}
 \country{USA}
}

\author{Benjamin Olsen}
\authornote{Work conducted while employed at Microsoft.}
\affiliation{
 \institution{Independent}
 \city{Redmond}
 \state{Washington}
 \country{USA}
}

\author{Weishung Liu}
\author{E. Glen Weyl}
\affiliation{
 \institution{Microsoft Research}
 \city{Redmond}
 \state{Washington}
 \country{USA}
}

\renewcommand{\shortauthors}{Working Paper In Progress. Lutz et al. 2025.}

\begin{abstract}
Previous work has found a lack of research in HCI on religion, partly driven by misunderstandings of values and practices between religious and technical communities. To bridge this divide in an empirically rigorous way, we conducted an interview study with 48 religious people and/or experts from 11 faiths, we document how religious people experience, understand, and imagine technologies. We show religious stakeholders find non-neutral secular  embeddings in technologies and the firms and people that design them, and how these manifest in unintended harms for religious and nonreligious users. Our findings reveal how users navigate technoreligious practices with religiously informed mental models and what they desire from technologies. Informed by this, we distill six design values–wonder, humility, space, embodiedness, community, and eternity–to guide technologists in considering and leveraging religion as an additional, valid sociocultural resource when designing for a holistic user. We further spell out directions for future research. 
\end{abstract}


\received{20 February 2007}
\received[revised]{12 March 2009}
\received[accepted]{5 June 2009}

\maketitle

\section{Introduction}
A majority of the world $\sim76\%$ is religious or identifies with a religion \cite{fahmy_how_2025}, but religious identity and wisdom has been largely left out of conversations and research regarding the design, development, deployment, and evaluation of digital technologies \cite{buie_spirituality_2013, wolf_still_2024, rifat_integrating_2022}. As digital technologies have become more ubiquitous, personalized and agentic \cite{arazy2015personalityzation, borghoff2025human, murugesan2025rise}, it is imperative to consider the whole person in such designs. And for many people, religion is a key, if not the key, component of their personhood \cite{fahmy_how_2025} and how they structure their lives and make sense of the world – including the technological one \cite{baucal_religion_2013, tripodi_searching_2018, rifat_integrating_2022}. 

Religion is one of the oldest human identities to be formalized in institutional categories and sociotechnical infrastructures – pre-dating and outlasting most national identities, racial schema and even many conceptions of gender. Religious institutions have contributed to several positive changes in the third sector – from education to assisting with refugees and  being loci of  peace building across the world. But, like many organized institutions that hold power and resources, nearly all major religions have also brought significant harms– from fundamentalists enacting violence at community and state actor level to reifying and being used to justify exclusion, discrimination, and harm to other (often marginalized) identities. 

But religion appears to be here to stay and is becoming an increasingly important global actor as population grows more quickly in the most religious regions and religiously-grounded political movements increasingly influence policy on matters including technology \cite{fahmy_how_2025, wormald_future_2015}. And while previous work has ascribed the paucity of religious research in HCI to its “taboo” and “complicated” nature \cite{buie_spirituality_2013, wolf_still_2024}, examples from Asia and other parts of the world show this may be possible to overcome \cite{ecklund2019secularityFINAL, eck2007prospects}. As with other sociotechnical systems, overlooking key sociocultural contexts risks exclusion and unintended harms — both by enabling bad actors in understudied environments and by neglecting communities whose needs and experiences are not examined.

This work situates religion as an additional, valid, and deeply human lens of experience, expertise, ways of knowing, and ethical frameworks and values. We see, and implore the broader HCI community to see, religion as a sociocultural and infrastructural resource to draw upon and carefully consider in the development and evaluation of technologies – a vision that may to some readers seem like fantasy but what many scholars have found to be an intertwining and ongoing dialogue in many parts of the world \cite{ecklund2019secularityFINAL, eck2007prospects}.  

But how does one approach the vast world of religion as it pertains to the equally large interactive technologies? What are some key considerations, concerns, and desires from religious users that technologists should take into account? Are these mental models and insights from religion that can help us approach technology design and evaluation writ large? And how can we distill such learnings into useful and actionable design values and recommendations to make approaching this intersection tenable for technologists? 

Our study takes an important empirical step in systematically documenting the intersection of religion and technology to inform technology design and practice. We conducted a landscape interview study with 48 religious individuals and experts across 11 faith traditions to address the following research questions:
\begin{itemize}
    \item RQ1: How do religious people conceptualize and experience technology, technologists, and technology firms? 
    \item RQ2: How does this, if at all, impact how they use technology in their daily (including religious) lives?
    \item RQ3: What do religious people desire from technology? 
\end{itemize}

In answering these questions through our work, we contribute:
\begin{itemize}
    \item Empirical understandings of how religious people conceptualize and experience technology and the institutions behind technology, and how this shapes their relationships to and desires from technologies. 
    \item Actionable design values for technologists seeking to build bridges to religious communities and integrate religion as an additional consideration in their work
    \item An agenda for cultivating more religiously literate technologies and key areas of future work for communities of practice within the HCI community. 
\end{itemize}

We hope this work serves as a springboard to empower technologists and HCI researchers to contend with religion as an additional sociocultural resource in their work.

\subsection{Key underpinnings and language of this work}\label{1.1}
We focus on  “religion” as opposed to "religious and spirituality" language \cite{zinnbauer2015religion}. This extends to our use of terms like “technoreligious practice” (where technology is incorporated in a religious practice). This is not to discount spiritual traditions, but to focus on the social organizations and histories of religion, in contrast to the more exclusively inward focus of the term “spirituality”. 

We adopt a sociological and sociocultural definition of religion. Sociologically, religion is a system of beliefs and practices that shape how people perceive the world, particularly the divine – and is manifested in social institutions \cite{cipriani2017sociology, roberts2015religion}. Socioculturally, religion is heavily integrated into other facets of culture via practices, norms, and symbols \cite{zittoun_sociocultural_2019, baucal_religion_2013}. Religious identities are in turn developed in context of cultural ones and these symbols and mental underpinnings form the basis for how people approach the world. Through these definitions, then, we see how religion is an infrastructure in the world but also a sociocultural resource that people access as they navigate the world \cite{baucal_religion_2013}.

Furthermore, consistent with this sociological emphasis, our study focuses primarily on “major faiths”, which encompasses a mixture of the largest religions (Christianity, Islam, Hinduism, Buddhism) and other organized religions prevalent in academic literature and policy circles such as Mormonism, Sikhism, Judaism, Shinto, and Taoism \cite{fahmy_how_2025}. We denote the 11 faiths represented in our empirical work in Section \ref{methods} and where our sample has limitations in Section \ref{limitations}.

When discussing “technology,” we focus on digital interactive technologies central to HCI (e.g. social media platforms, search interfaces, VR, AI chatbots, etc). Our research takes a sociotechnical approach: religion and other social values both shape and are shaped by technology, and system design must account for both the technical and social. 

This work draws from interpretivist epistemologies \cite{alharahsheh2020review}, centering lived experiences and expertise of religious stakeholders. \textbf{We do not seek to promote religion over other epistemologies but to systematically study it as a sociocultural resource relevant to technologists and researchers making and studying technologies that interface with a holistic person’s lived experiences.}

\section{Related Work}\label{related_work}
We first document and situate our work within religion's position as a sociocultural resource by other fields, namely sociology and religious studies (\ref{2.1}). We then discuss technoreligious practices and how religion has (and hasn't) been covered in HCI research (\ref{2.2}). Finally, we consider how other sociocultural identities and frameworks have been brought into HCI (\ref{2.3}). 

\subsection{Religion in society and technological innovation}\label{2.1}
Arguably religion is one of the oldest, and longest studied, human institutions. Along with taking a sociological and sociocultural definitioning of religion in our work (see \ref{1.1}), we look to studies of how religion has intersected with cultural institutions like education \cite{grace2004making, kohlbrenner1961religion}, healthcare \cite{pesut2008conceptualising}, and even science \cite{ecklund2019secularityFINAL}. Such sectors and empirical work has shown how deeply tied religion is for many communities to care work, such as mutual aid and assistance in times of crisis \cite{vu2025exploring, wilson2016refugee, speers2023caring} but also individual mental well-being \cite{green2010religion, koenig2001religion, cohen2017relation}.

Studying religion requires “religiously literacy”, which is akin to many literacies the HCI community considers, such as digital \cite{ogbonnaya-ogburu_towards_2019}, AI \cite{cao_empowering_2025}, and media literacies \cite{bulger_promises_2018}. Religious literacy is the competency to understand religion’s diversity and influence and recognize it as a significant human phenomenon while being able to analyze and engage with its impacts \cite{chan_religious_2021, hannam_religious_2020}. We hope this literacy could be added to technologists’ mental toolboxes as an additional (albeit not superior) lens, falling in line with operations research about benefits of religious literacy in the workplace \cite{grim2016socio, grim2015modern}. 

The diversity of religions is a key aspect of any work in religion. We approach this work largely from a “plurifaith” perspective, drawing from a history of multi and interfaith dialogues across time and sectors, as well as Eck's idea of pluralism as rooted in religious case studies \cite{usman2023does, ariarajah2019interfaith, eck2007prospects}. Now, such pluralistic dialogues are expanding to technology, such as with the “Rome Call for AI”, a multifaith effort led by the Vatican on AI Ethics \cite{zengarini_world_2025} and other similar efforts \cite{Elmahjub2023IslamicEthics, RFBF_AI_Faith_Ethics, Templeton_PerlisPromises_AI, AIandFaith_ReligiousEthics_AgeAI}.  

This notion of religious dialogue about and in conversation with technology may be surprising to some readers. Religion is often held, especially in the Anglosphere, as dichotomous with science and technology, which in turn are associated with ideas of secularism intended to exclude religious influence. But, like many things – the reality is much more messy, and human. Ecklund et al, in their landscape study of hundreds of scientists, found that religion does not need to be in conflict with science and technology, but is a massively important cultural resource in conducting scientific work \cite{ecklund2019secularityFINAL}, a position much more recognized in Asian cultures, but also accepted in many non-English-speaking Western societies \cite{grim2015modern, clart2003religion, usman2023does}.

To this end, we also look towards work in religious literacies (and other literacies) that seek to enable technologists and designers to engage in informed ways to understand how religion impacts people (both individually and in groups) and how they receive technologies.

This framing of religious literacy, plurifaith dialogue, and a sociocultural and sociological approach to religion is imperative to our study’s goal of uplifting religion as an additional, but not superior, consideration in technology. 

\subsection{Technoreligious practices and research in HCI and religion}\label{2.2}
The complementarity between religion and technology is a small and understudied subfield within HCI \cite{wolf_still_2024}. Although there have been more community calls to examine this intersection \cite{kitson_diving_2025, rifat_integrating_2022}, systematic literature reviews in 2013 and 2024 have pointed out a paucity of papers deeply engaging with religion and religious people \cite{buie_spirituality_2013, wolf_still_2024}. These reviews emphasize that the majority of work in HCI engages with religion as a peripheral topic (i.e. its relationship to grief) or via design activities and case studies \cite{wolf_spirituality_2022, caidi_recapturing_2023, claisse_keeping_2023, halperin_miracle_2023, gaver_prayer_2010, smith_i_2020, wyche_sacred_2009}. These contributions are valuable in designing technologies and learning from technoreligious practices. But they do not incorporate religious frames as central theoretical and empirical focuses for how people live their lives and interact with technology, which is a smaller space of work \cite{wyche_extraordinary_2009, wyche_technology_2006, ames_worship_2015, oleary_examining_2021, bell_no_2006}. 

Much of HCI's work on religion is around technoreligious practices -- where religious practices are enabled or augmented by technology, which can in turn inform technology design writ large \cite{wolf_spirituality_2022, caidi_recapturing_2023, claisse_keeping_2023, halperin_miracle_2023, gaver_prayer_2010, smith_i_2020, wyche_sacred_2009}. Technoreligious practices predate today’s digital interactive technologies – religions have always adopted technology from spiritual and religious calculator tools in ancient civilizations to the revolution of the printing press and its role by and with religion \cite{stahl2009god, neugebauer1945history}. Technologists and designers, meanwhile, have sought inspiration from religious tenants and communities in how technology and online spaces can be more affirming for religious people and others \cite{halperin_miracle_2023, smith_what_2021, smith_sacred_2022, premazzi_religious_2015}. 

The “AI era” is changing some of these relationships and technoreligious practices. Increasingly, religions are adopting AI tooling in technoreligious practices, such as AI mediated religious content across a variety of faiths \cite{Brumfiel2025AIBible, IslamAndAI} or even AI confessionals \cite{Katz2024AIJesusConfessional}. In parts of the world, such as in India, large sectors of the startup economy, often empowered by AI, are directly focusing on astrology and religion \cite{the_economist_astrology_2025, Zaffar2025}. Users are increasingly turning to AI agents with religious and existential questions. And while there are some fine-tuned, specialized models are run by religious organizations for this type of use \cite{IslamAndAI}, many mainstream AI tools (i.e. ChatGPT) fall short in this dimension. Cases of psychological harms from chatbots with religious undertones, from spiritual psychosis \cite{tangermann_chatgpt_2025}  to chatbots providing self mutilation instructions associated with Satanic rituals \cite{shroff_chatgpt_2025} are becoming increasingly important. Scholarship highlights flaws in general models’ ability to understand religiously contextualized image-based abuse of women and children in religious clothing \cite{waheed_for_2025}, and some theologians caution AI may invite dark forces into the world \cite{rossetti_exorcist_2021}.

Regardless of one’s beliefs there is ample space for religiously informed investigation in evolving interactive technologies. Our work seeks to extend the niche of religion in HCI by focusing on religion as a foundational guiding force to inform design and computation at large.

\subsection{Sociocultural approaches to HCI}\label{2.3}
There is much work in HCI about how human identity is encoded in and impacted by technological systems. Sociotechnical systems often compress (via over-simplification or sometimes harmful umbrellaing), under-represent, or even preserve harmful and dehumanizing understandings of human identities \cite{lutz_were_2024, das_toward_2023, ghosh_i_2024}. This identity based work suggests that it will never be fully possible to digitize the messy, multifaceted experience of being human. 

To help protect communities from attacks targeting sociocultural identities, researchers and practitioners have developed sociotechnical frameworks that guide the design and deployment of more inclusive and resilient technologies. Such frameworks have encompassed stakeholder concerns, such as trauma-informed computing or conflict-sensitive design \cite{chen_trauma-informed_2022, otuu_how_2025, scott_trauma-informed_2023}. Other frameworks have centered social identities, from feminism and critical race theory to geo-cultural approaches addressing computing for and with the Global South, often taking an intersectional approach in the style of Crenshaw \cite{crenshaw2013demarginalizing}, focusing on axes of marginalization and power in sociotechnical systems \cite{bardzell_feminist_2010, ogbonnaya2020critical}. More generally, value sensitive design has focused on ways to distill the often complex systems of human values into technology design from the outset, particularly through stakeholder engagement \cite{friedman_value-sensitive_1996} – and this framework has been applied to a variety of case studies, including aforementioned ones and in design frameworks like Design Justice, which reckons with how marginalized stakeholders may influence design \cite{costanza2020design}. 

We seek to extend this work by bringing religion as an additional sociocultural resource into computing. We particularly draw from Friedman’s value sensitive design \cite{friedman_value-sensitive_1996}, seeking to inform design values as non-prescriptive, guiding directions from religion. In building these values, we focus on religion as an intersectional identity in conversation with other social ones (i.e. race, culture, language) and viewing religion as intersecting social spheres and institutions that influence how people make sense of the world \cite{simmel_sociology_1908, crenshaw2013demarginalizing}.

\section{Methods}\label{methods}
\subsection{Interview design and recruitment}
This study used semi-structured and narrative-focused interviews, following approaches used in prior work on how personal identities intersect with technology \cite{lutz_were_2024, simpson_for_2021}. As is typical in semi-structured interviews, protocols served as a guide—some questions were skipped, new ones emerged, and participants could always decline to answer \cite{kallio2016systematic}.

Our interviews began by asking how religion appeared in participants’ lives—through their own beliefs or views of religion as a system. All participants disclosed their personal beliefs. We then asked how technology appeared in their work, faith, and daily life, as well as their views on secularism and religious practices. This opened discussion of if religious and technological institutions had relationships and what each could learn from the other. Finally, participants were asked what they desired e from technologies and their creators, and invited to share any additional reflections.

48 interviews were conducted by the first author on Teams or in person, lasting an average of 113 minutes with a minimum of 47 and maximum of 200 minutes. Most interviews were ~90 minutes. 

Participants were recruited via email. Many were cold emails (n = 20) – such as emailing to faith leaders, religious scholars, and individuals working at the intersection of religion and technology. Others came from professional and faith groups the team had access to (discussed in 3.3) (n = 22) or from the first author’s academic networks (n = 6). 

For this study, our goal was to capture opinions from a broad tent of plurifaith stakeholders – religious leaders, religious studies or technology ethics scholars, technology policy analysts and ethicists, and people working at the intersection of technology and religion (see Table \ref{tab:participants}).

\begin{table}[H]
\centering
\begin{tabular}{|l|p{5cm}|l|p{5cm}|}
\hline
\textbf{PID} & \textbf{Participant Description} & \textbf{PID} & \textbf{Participant Description} \\ \hline
P1 & Community organizer; Diné\footnote{Diné are an indigenous tribe in the United States, commonly called the Navajo} Religion & P2 & Imam; Muslim \\ \hline
P3 & Nonprofit director; Mormon & P4 & Research institute director; Catholic \\ \hline
P5 & Professor, political science; Christian & P6 & Nonprofit leader; Catholic \\ \hline
P7 & Researcher of online populism; Jewish& P8 & Priest; Catholic \\ \hline
P9 & Journalist; Christian & P10 & Chaplain and author; Humanist \\ \hline
P11 & Tech worker and activist; Muslim& P12 & Professor of Jewish Studies; Jewish \\ \hline
P13 & NGO leader, technologist; Taoism & P14 & Monk and scholar; Buddhist \\ \hline
P15 & Interfaith leader/technologist; Bahá'í & P16 & Faith organizer/tech worker; Pagan \\ \hline
P17 & Professor; Christian & P18 & Pastor and community organizer; Christian \\ \hline
P19 & Policy analyst; Christian& P20 & Mosque outreach lead; Muslim \\ \hline
P21 & Policy analyst; Evangelical Christian& P22 & Genocide and theological scholar; Buddhist \\ \hline
P23 & AI CV Scientist; Sikh & P24 & Religious studies (Hinduism) professor; Not religious \\ \hline
P25 & VC investor; Catholic & P26 & Religious tech researcher; Jewish \\ \hline
P27 & Policy analyst; Christian& P28 & Rabbi; Conservative \\ \hline
P29 & Lutheran pastor & P30 & Children’s program director; Lutheran \\ \hline
P31 & Ministry/nonprofit leader; Evangelical & P32 & Nonprofit founder, chaplain; Sikh \\ \hline
P33 & Mental health counselor, imam & P34 & Hindu priest \\ \hline
P35 & Familial steward; Shinto shrine& P36 & Religion and work scholar; Greek Orthodox \\ \hline
P37 & Nonprofit leader; Christian (Church of Jesus Christ of Latter-day Saints) & P38 & Policy analyst; Christian\\ \hline
P39 & VC investor; Christian (Church of Jesus Christ of Latter-day Saints) & P40 & Innovation technology researcher/volunteer priest; Hindu \\ \hline
P41 & Rabbi; Liberal & P42 & VC investor; Catholic\\ \hline
P43 & Law Professor; Jewish & P44 & Episcopal priest; former mental health counselor; Christian \\ \hline
P45 & AI Safety researcher; Buddhist & P46 & AI scholar; Muslim \\ \hline
P47 & Ethics and policy researcher; Christian & P48 & Rabbi: Reform; Jewish \\ \hline
\end{tabular}
\caption{Participant IDs and Descriptions}
\label{tab:participants}
\end{table}

\subsection{Interview analysis}
Most interviews (n = 45) were transcribed, but all were taken extensive notes of. The analysis was primarily performed by the first author, in a grounded theory style leading to a thematic analysis \cite{braun_using_2006, charmaz_constructing_2006, chapman2015qualitative}.

Following the approach of Charmaz \cite{charmaz_constructing_2006}, transcripts and notes were first open coded by the first author. Each interview was then memoed into a summary. Each week, the author constructed a memo across that week’s interviews, axially coding across the open codes from each. This memo was presented to the research supervisors and discussed as a research team. The author then applied a second round of iterative coding to all interviews based on that discussion and the week’s memo, testing mini hypotheses and observations from the first round of analysis and across all interviews, as is typical in grounded theory. This process continued weekly over the course of seven active interview weeks. The first author aimed to schedule approximately six interviews a week, to not overload analysis and result in interview fatigue. After all interviews were complete, memos were compared and coded across all 48 interviews, with some select interviews being coded a third time. Across this comparison and memoing, themes were identified and extracted to sturcture findings \cite{braun_using_2006}. Main themes around secularity, how the relationship of religion and technology has changed over time, religious technology use and boundaries, and religious desires for technology emerged as most prevalent, and are presented in our findings.

\subsection{Ethical considerations and positionality}
This study took place in a large U.S. industrial lab and was approved by the company’s IRB which oversaw our consent forms, protocols, and recruitment materials. In email recruitment, participants were told about the study goals. They signed electronic consent forms, which they were able to keep. At the start of each interview, participants were reminded of consent and asked for permission to record and transcribe. The first author sent a participant pre-print of this study to participants for them to understand how their data was used and to request quote and identifier description changes. 

This study was led by a PhD student research intern, with full-time employees serving as supervisors and mentors. Two supervisors had ties to religious policy and professional organizations, which the intern was introduced to, and these connections formed a major part of the interview pool (n = 22). This team represents individuals with experiences or active ascriptions to Jewish, Catholic, Buddhist, and other spiritual and agnostic traditions. The PhD student comes from an HCI program and the supervisors have academic backgrounds in Religious Studies, Economics, Business, and Philosophy. No one on our team identifies as an atheist and we acknowledge we hold generally positive views of religion as a resource for many people, which colors our findings. To account for this, we engage with literature and interviews with scholars that hold more critical views of religion and we approach this work committed to religion as an additive, and not superior, lens and aligned with the UN's definition of religious freedom that supports every person to practice or not practice a religion of their choice \cite{UN_UDHR_1948}. 

Given the human subjects nature of this work, we disclose the first author's positionally, which was also offered to participants. The first author and interviewer was raised Catholic in a deeply diverse American Southwest community and was exposed to several faith communities. However, they are presently identify as agnostic and spiritual. As such, they consider and have empirically seen religion as a deeply sociocultural phenomenon that intersects with several human identities. We found in this study, given its breadth and engagement with so many faiths, that having an interviewee with a religious background but no particular affiliation was helpful for building rapport and honest reflections about religious specificities. 

\section{Findings}
First, we summarize how participants conceptualize secularist embeddings in technologists and technologies (\ref{4.1}) and how they feel this has resulted in religious exclusion, leading to unintended harms (\ref{4.2}). Secondly, we summarize how religious epistemologies inform technoreligious practices (\ref{4.3}) and desires religious people have of technology (\ref{4.4}). 

\subsection{Secular embeddings and assumptions of technologists and firms influence technology design, development, and use}\label{4.1}
Technology firms and technologists are seen as “secular” by participants. Participants conceptualize secularism as a non-neutral stance used by institutions to distance themselves from what they perceive as embedded assumptions and biases. They described firms’ commitments as “dogmatic” (n = 17) and “religious in a sense” (P10), likening Silicon Valley to a religion-like system grounded in the belief that technology can solve all problems and that more is always better. Participants viewed power and profit as the dogmatic forces driving such beliefs. It is worth noting, a minority of religious scholars (n = 4) explicitly felt religious institutions were deeply power motivated and many religions had abused their power in ways similar to Silicon Valley. However, all participants felt science, technology, and, in turn secularism are were perceived as more "neutral" than religion, but participants saw neither sphere as neutral. Rather, religious institutions were seen as more transparent about their assumptions and beliefs and did not claim neutrality. Participants hoped technology firms might learn this from religion, and acknowledge their values and commitments. They view this as a first step to challenging the notion that technology must advance at all costs, regardless of labor and environmental concerns, of which participants had many. Unacknowledged values, they argued, shape not only design, development, and deployment but also how technology is communicated, making it hard for communities to make informed decisions. 

\textit{“It [a company] can be secular in the sense that it doesn't outwardly embrace a specific organized religion, but the idea that it's actually neutral, that it's not based on some substantive moral commitment that it's just a procedural entity – I think that's…a helpful illusion, sort of noble lie if you want. But I don't think it's real and it's become increasingly exposed as you can't really maintain that.” – P9, Journalist, Christian}

Participants do not wish for religion to usurp non religious ideas in technology firms. Instead, they pointed to Eastern perspectives, where technology and religion are often more integrated, with religious case studies and sociocultural norms—such as ethical principles—informing practice, something P40 encountered in his study of Indian startups:

\textit{“I didn't go into conversations with it thinking they are religious startups, but…religion naturally came in the conversation...[about]...what they do and why they do it...Venturing that is in some way inspired by religion, right? In the Indian setting, I think it's it's sort of organic…a VR startup in their employee handbook, they're setting guardrails around technology inspired from the scriptures.” - P40, Technology scholar, Hindu}

\subsection{Secularism leaves religion behind and out of technology}\label{4.2}
Participants felt technology firms have (perhaps unintentionally) largely excluded religion, and this history shapes their relationship to technology and unintended technologically mediated harms to their communities. Participants desire to have religious stakeholders included, but they are pessimistic and therefore unclear what inclusion may look like. 

\subsubsection{Exclusion has led to a rift between religion and modern technology}\label{4.2.1}
Participants feel that religious communities have been increasingly excluded from the “cutting edge” of technology over time, with early stages of communications technologies generally empowering them, marginalization beginning in during the social media era and great trepidation regarding the expected AI future. 

Participants generally viewed religion’s historic relationship with the early stages of digital technology, such as early computing, telecommunications, and early social media as crucial for maintaining connections and increasing access to religious content. This era also marked the rise of “big data,” with digitized, searchable holy texts available across languages, something they valued greatly. Peer-to-peer digital communities helped digitize and keep religious communities in touch, especially for physically distributed faith communities (e.g. Muslims in the West).

\textit{“Mormon.org was by far the most sort of progressive, advanced church tech initiative of its time….It had a massive community element where everybody in the church was instructed to create their own mormon.org profile…in 2008, that was crazy unheard of and it was very successful…you could find somebody who was local to you…and [help with] member community building.” - P3, Nonprofit leader, Mormon}

Attitudes in participants’ faith communities shift, however, in the era of personalized social media around 2010. Here, shifting from human curation and building of intentional digital space to hyper-commercialization and algorithmic recommendations is seen as undermining community. Participants connect this moment to more broad political and social polarization in and outside of religious communities. As P29, a pastor, said, \textit{“No one wants politics in church, but the algorithm sure does.”}. In their communities, participants see content targeting religious views and strengthening internal divides. They also see this wave as increasing discrimination (via anti-religious online content) and even persecution of religious groups, such as the Rohingya genocide and Facebook’s role in it\footnote{This example was often mentioned by participants, and is very well documented in literature \cite{mozur2018genocide}}. Overall, this era is seen as fueling isolation via manipulative technology and content.

\textit{“There’s so much misinformation there…da'wah is not meant to be digital propaganda, it’s…personal invitation and relationships.” - P2, Imam [da’wah is the inviting of people to Islam]}

\textit{"The way that polarization has been fueled by technology, 40\% of Christian pastors say they've thought about quitting in the last 5-6 years. And a lot of if you ask them why the top three reasons are anxiety, isolation, burnout. And then when you ask them why are they isolated and burned out, polarization is number one. So it's it's everywhere. I mean, it's coming on all sides.” - P31, Nonprofit leader, Christian}

\textit{I worry about antisemitic content and conspiracies…but also content that tries to make us turn our backs on our siblings from other religions...that is when we will fail, if we do not have solidarity across people of faith...That is when religious freedom is under threat. - P41, Rabbi}

Most participants saw religious communities as a final redoubt of rich community life, threatened by digital encroachment. In contrast, a minority of participants from primarily Eastern, polytheistic, and/or diasporic backgrounds saw technology as a site for creative experimentation, including VR and gaming content delivering religious experiences. P35, whose family maintains a Shinto shrine in their community, discussed how immersive projections revitalized the temple and helped with maintenance. These stakeholders saw technoreligious practices and innovation as a natural step for their religions’ modernizations, but still an community by community choice.

\textit{“Religious communities have weathered the storm a bit…because of that they are important places for people to develop ideas of what is right and wrong…and develop their approaches in this community” - P26, Technology researcher, Jewish}

\textit{“I don’t know how much longer church can compete with cellphones and games…[they're so] addictive.” - P30, Children's  programming director, Lutheran} 

\textit{“It started as an experiment. I was unsure but I think [the] spirits welcomed the new energy…[and now] the young come more. - P35, Shrine Steward, Shinto }

Participants perceive the current and evolving AI-driven wave of technology as potentially amplifying harms. They also perceive AI as a potentially more disruptive force than social media and a conceptual challenge to doctrines about human nature. These high stakes motivate increased focus among stakeholders in shifting technology towards pro-religious and pro-community directions and away from the harms they see arising from algorithmic curation and top-down Silicon Valley governance. They were also concerned that religion and technological advancement occurred at different, and contradicting, speeds -- which would leave religious people again excluded in this AI wave. They feel solidarity in this concern with other marginalized (not-necessarily-religious) communities (i.e. people of color, workers displaced by automation). 

\textit{“The speed of technology development is just ‘because we can, we should, and we should faster than the other guy’. No reflection, no debate. And religious people care about that. We take time to decide…ask scholars…reflect. And I think…[that] is part of the reason we are not at the table.” - P33, Mental health counselor, Imam}

\textit{“I worry about working people. Not just religious ones. It’s like the information environment is against people…Now the physical environment [with] data centers and jobs being cut?” - P23, Computer vision scientist, Sikh}

And, although there was a consistent distinction between Eastern and Western faiths, with the former more optimistic of technology, all shared concerns that technology’s rapid development was in tension with religion’s more intentional pace – which may exclude the latter.

\subsubsection{Exclusion causes harm, but an inclusive vision is unclear}
Participants felt exclusion began upstream of development through cultural artifacts and professional norms that belittle religion, as evidenced in conversations around secularism and neutrality. These norms were not only stinging to participants, but they theorized were reified in both technologists' and massive datasets' biases. 

\textit{"[talking about anti-religious TV content] I heard the contempt, you know? And yeah, I mean, I'm concerned about that...I've got a lot of my own data on the Internet, lots of sermons...but also, if you want to find out who we really are: you can. And if you don't, if you would rather just think we're idiots, well, I don't know if there's anything I can do.” - P44, Pastor, former mental health counselor, Christian}

\textit{“With so much hate against Muslims online and AI now deciding stuff like jobs…I worry that AI could be the next post-9/11 era for discrimination.” - P20, Mosque outreach leader, Muslim}

Participants, particularly non-white participants, saw religious bias as similar to and interlinked with other biases. Meanwhile, white participants often saw anti-religious bias as a last frontier of bigotry. But all participants see anti-religious bias as an additional form of discrimination not actively being tested for in systems. A lack of testing, participants worried, has manifested and can manifest in further  real-world harms—such as image-based abuse beyond Western standards of nudity and concerns about hateful content generated by AI. Participants want religious stakeholders at the table to raise such edge cases and normalize religion as an important dimension of safety considerations. 

\textit{“Taking faith seriously means that someone has to respect that being depicted without a hijab for me is trauma. It is not fashion....They do not have to understand it but they have to respect it. [If] a quarter of the world is Muslim they need to codify that it is image-based abuse that is happening a lot more than deepfake, hardcore porn. How is it still happening? Obviously they aren’t testing enough.”  - P11, Activist and tech worker, Muslim}

Participants lacked a clear vision of what inclusion may look like. Many suggested formal involvement of religious leaders on ethics committees and advisory bodies to technology companies. Others suggested stronger commitments to religious freedoms in the workplace to have technologists become more attuned to religion. But most visions were vague and shrouded in caution. Participants feared some religions (or opportunistic actors within them) might gain disproportionate influence. But, more prominently, participants were greatly pessimistic technology companies would take religion seriously and that technological mediated harms could be solved by tech companies. They worried religion could become the next superficial corporate promise, wasting stakeholder time when many deeply rooted issues technology exacerbates required policy and offline, communal solutions. Participants instead had more hope for technology regulation efforts, pointing to recent laws around age verification and protection of one's likeness. They hoped these efforts would engage a variety of cultural stakeholders, including religious ones, rather than leaving it up to companies. However, participants stressed that companies needed to contend with these matters, as people were imbuing meaning onto technology and that could cause harm and backlash.

\textit{“I just think the problem is these companies, they're so entrenched and enmeshed with this economic incentives and sort of way of conceiving of the world, I don't think there will be any change until they are restrained by law and forced to change.” - P38, Policy analyst, Christian}

\textit{“[Anonymous Company] is not a religion, it’s a company that makes money and technology…that’s what it does. But understanding...that people are going to look at technology as a God and a source of meaning....[How] do you want to set yourself up for pitchforks that will come when people realize their God is fake? Because we’re there.” - P7, Researcher, Jewish}

\subsection{Religion as a resource that shapes technoreligious practices}\label{4.3}
Our participants describe religion as a central identity that for them is more core and inherent than their other identities, including race, gender, ethnicity, and what they do for work. Religion can help them navigate other identities, as P18 puts it: 

\textit{“My peace comes from God. It doesn't come from being a Black man in America.” - P18, Environmental activist, pastor, Christian}

Participants conceptualize religion as a resource they tap into to guide their lives, including technology use. For most participants, technoreligious practices often encompass information access or administrative tasks. These and other practices are most commonly as “extending” practices or “archiving” religious knowledge. Extension was by far the most common, especially when individuals were on the go, such as finding a temple when on vacation, using a prayer app for timed prayers, or being able to access scriptures on their phones. For some religions, such as Mormonism and Catholicism, these technologies come from in-house teams making suites of apps for their congregants. For most, technologies of unknown provenance or grass roots efforts were popularized for religious activities. This included Muslims pointing out the prayer time app as one of the most popular apps in the world or a variety of online guides for religious diets (i.e., halal, kosher), such as P28, a rabbi had used:

\textit{“It [Kosher food website] gets the job done…I think a rabbi’s son made it. I’m not sure. Muslims have a nicer website for halal food, though.” -P28, Rabbi}

Some, particularly of Eastern faith traditions, take a more active and creative role in this extension. Hindu stakeholders described several VR experiences and custom apps they have seen. Buddhist and Shinto stakeholders described interactive temple guides made from customized AI agents, with a Buddhist monk (P14) discussing how he had customized a variety of AI workflows to bring him closer to Buddha and help in his meditation practice and teaching:

\textit{"In the process to become Buddha you need to understand your human nature at first, and then you need to, you know, reduce your humanness… I really think and even experience that we can use AI, in particular generative AI in large language model like a functional Buddha who can help me understand my own humanness." - P14, Monk, Buddhist}

Many religious individuals set firm boundaries around technology, including outright rejection. A conceptualized separation of “work” vs “Spiritual Work” facilitated many of these boundaries. These included examples of reserving prayer and sermon writing for humans (Spiritual work), but not seeing issue with tech use in upstream research or with technology helping automate logistical tasks (work). But participants acknowledged these boundaries were contextual, pointing to religious practices going remote during COVID-19. In general, our participants preferred the in-person experiences over digital ones, but felt that technological choices were valid and highly personal decisions between a person, the divine, and their faith leader.

\textit{“God moves through us...the message that you get from God has to come from out of you, not the machine…it gives me a very icky sense to think that someone’s going to preach an AI’s sermon..creativity comes from something else…the spirit moves in a way…will that happen with AI? And if it can’t, how do we continue to tell people to do the harder spiritual or creative work?” - P5, Political scientist and former pastor, Christian}

Broadly, participants aimed to align technology use with their religious values. Concerns like dependence and addiction were seen as conflicting with these values, sometimes framed as a form of idolatry. Even in faiths without the concept of idolatry, participants stressed machine connections shouldn't undermine free will through manipulation or addiction. Across traditions, participants recognized that religion has long negotiated its relationship with technology—and believed this negotiation would continue.

\textit{"It requires is to get out of this passive mode. It doesn't mean that you can't use a tool, but what it does mean you have to evaluate whether any particular tool is actually responsive to the needs of your community…it takes effort….You have to create…these mechanisms of reflection...Wat was it doing? Is it enhancing agency? Is it reducing agency? Is it short cutting processes like relationship building?” - P15,  Interfaith leader and technologist, Bahá'í}

\subsection{Religious desires from technology and technologists}\label{4.4}
Although our participants tended to maintain strict boundaries with technology—often refusing it or being “medium adopters” (P44)—they still expressed desires for technology design beyond  treating religion as a valid testing ground. In our discussion, we distill these desires into actionable design values for technologists (\ref{5.1}). 

\subsubsection{Preserving relationships and communities}
Religion grants participants a strong commitment to community and relationships, prioritizing relationships between themselves, the divine, and their community over individualism. In a world of chronic loneliness, they experience religious spaces as thriving counterexamples, and feel fiercely protective over them. They found AI, particularly with its current marketing as a “companion” or “assistant” threatening to human-human and human-divine relationships. The addictive nature of many interactive technologies (i.e. social media, video games, and chatbots) is also seen as an erosive threat. This also resonated in Eastern and more animistic faiths, who were more comfortable with AIs posed as entities (and multiple notions of nonhuman entities at large), but who stressed the importance of community centric AIs over  individualized "servants". 

\textit{“Both Taoism and Shinto has a very animist idea...we are very comfortable with spirits. Either ancestral spirits, steward spirits, or spirits of some nature feature that has a contained moral scope...so a steward of that commons cares only really about that commons and the relation around that commons, which is very different from the Abrahamic Tradition. So… thinking of [AI] as a kind of purely loyal servant to…human…to me, it always sounded very, very strange...Whereas a relational spirit that cares about a commons [like] Wikipedia. That makes much more sense to me.” - P13, NGO leader, Taoist}

A key portion of this relationality was dedicated time to be away from frictionless, individualized technology experiences (i.e. social media) to rest or conduct Spiritual Work such as meditation or pilgrimages. Jewish and Christian stakeholders emphasize not only the importance of rest through the Sabbath, but also that the Sabbath was meant to be extended to others (even animals). Although Jewish communities have negotiated technology integration (i.e. Sabbath lamps) into this religious practice, they found much of modern tech's focus on constant availability and productivity to be a challenging, and perhaps harmful, norm. This was especially evident with the notion of AI relationships. Most religious people are not advocating turning off the grid once a week. Rather, they are challenging the notion of building technologies claiming to be “human-like” and giving them no limitations or rest – and fear what precedent this sets for human-human relationships.

\textit{“I say all that to say, and if that's the case, then we should give the oxen their Sabbath as well and that oxen are  these AI systems. Now what are the impacts of that?” - P36, Faith and work researcher, Greek Orthodox}

\textit{“Smaller scale examples like Sabbath observance I think is really important in Jewish and Christian traditions….it's not just like you observe the Sabbath, but it's you must give Sabbath to everyone who is underneath your power and purview. ” - P21, Policy analyst, Evangelical Christian}

Overall, participants wanted technologies that fostered human and community relationships over human to AI ones, especially ones with unrealistic expectations. 

\subsubsection{Embodiedness outside of technology}
For many, religion is an embodied experience that technology cannot replicate. There are cases where digitization is helpful, such as on-the-go scripture access or live-streamed services. But still these experiences felt short and participants found in-person, physical engagement as essential to their practices and communities. 

\textit{“Zoom does the job but you can tell this was made for meetings not worship.” - P29, Lutheran pastor}

\textit{“Every weekend, right, we have to try to get people to show up in person and have this slower, more human experience when they could say 'Well, I get my religion from this podcast and I do this other thing online'….[but] it's not the same thing.” - P31, Nonprofit leader, Evangelical Christian}

\textit{“Catholicism is always kind of trying to be old school…we are kind of the original touch grass kind of movement….we always had rather this idea of trying to use…natural things…it's not aversion to technology, it's more about that there is a different experience and ambience when you try to limit amount of technology…In interpersonal ministry…we try to avoid it or we don't need it because human contact is one-on-one, but of course we use calendars, we use e-mail services….what is I think is technology is awesome for the accessibility to people and content.” -  P8, Catholic Priest}

They emphasized spiritual experiences often transcend even written record, such as the importance of oral tradition and connections to place and nature — something they find difficult to datafy. This stands in contrast to what they perceive as a technologist’s tendency to treat written or datafied records (which many religions have plenty of) as the sole ground truth, often overlooking embodied, non-digital experiences. However, for traditions who were diasporic or, like indigenous ones, shrinking, digitization offers an important archiving ability, but how to accomplish this in a way that feels true to and honors these traditions feels difficult to participants. 

\textit{“If you are reading the Holy Quran, you can read it on a device or a book, but ultimately, the deepest connection a Muslim can create is in its memorization and recitation from memory. There is no technological replacement.” - P46, AI Scholar, Muslim}

\textit{“I am on a goose chase to record elders to preserve practices but I understand that is a knock off for the oral traditions we can no longer pursue. But I’m still grateful for it." - P1, Community organizer, Diné}

Although participants are split about how a machine should or should not summarize or interpret holy scripture, they all believe there is an intangible quality to religion that cannot be perfectly digitalized, and want technologists to acknowledge this limitation.

\subsubsection{Humility and avoiding idolatry}
Participants wanted technologies and the companies making them to demonstrate humility—a virtue across most faiths. They sought honesty about what tech can and can’t do, pushing back against hype, misleading terms of service, and the assumption that more tech is always better. While they recognized technology’s benefits, they emphasized it only helps when carefully deployed and accessible, and that technology is not the root solution for human problems.

Participants often (n = 31) invoked concepts of idolatry, or prioritizing something above the divine – even for participants without idolatry as a concept in their religion. Several participants (n = 17) connected this concern to the Tower of Babel\footnote{The Tower of Babel story tells how humanity, once united with a single language, sought to build a tower to reach the heavens, but God thwarted their prideful ambition by confusing their speech and scattering them across the earth (Genesis 11:1-9, New Revised Standard Version)}. P43 and P24 described the story not simply as a tale of hubris, but as a warning against valuing advancement (what they equated with technology) over human dignity. Participants felt this was a useful caution against building simply because one can, rather than because one should. Furthermore, Babel illustrated to participants how technology (the tower) could consolidate authority among its builders, enforce uniformity, and even attempt to rival God—a sacrilegious ambition. At the same time, participants acknowledged that today’s technologies, such as large language models, give rise to new “idols” by generating language and conversation in ways that further blur the line between human and machine authority.

\textit{“The golem didn’t have language...and that’s what defined it. So when we talk about AI, we’re talking about something that imitates language but lacks the soul behind it.” - P43, Law Professor, Jewish}

\textit{“Idols...have hands but can't hold anything, eyes but can't see, mouths but can't speak...when we worship an idol, we think that we have this new special insight, but in reality we're becoming blind and it's a thing that's actually disengaging us from the world, not giving a special power over it.” - P21, Policy analyst, Evangelical Christian}

\textit{“The golden calf couldn't have a whole conversation with you, right? I can call the AI agent an idol, but it has powers that the idols themselves didn't have. And those powers are based on…millions of hours of of labor of data labelers…”- P10, Chaplain and author, Humanist}

Idols demand singular focus and uniformity—parallels participants drew to the “AI race.” Through the lens of idolatry, they critiqued both the belief that AI superintelligence is desirable or possible and the view of AI as an all-knowing oracle, emphasizing that idols are neither holy nor human, but mere imitations of knowledge and holiness.

\subsubsection{Embracing uncertainty and wonder}
Religion allowed participants to be comfortable with uncertainty and not knowing. P5 described this as a “healthy dose of doubt” that allows believers to struggle with and interrogate their beliefs. For some, not knowing is an important part of their spiritual journey. This uncertainty, and comfort with it, was deeply related to holding multiple truths at once, even in tension. Participants felt that information spaces focused too much on absolute facts and summarizations, implying every question had a knowable answer. Leaving space for not knowing and actively embracing wonder as sacred and important stands in tension to things like automatic summarizations. Rather, they acknowledged that truth is more fluid in faith and that some knowledge is meant to be earned and not fully discrete. 

\textit{“What a relationship with Christ does is it holds you together and it allows you to thrive regardless of the situation you find yourself in. It allows you to have a hope when everything is lost. Allows you to see life, not just in the immediate now…it allows you to embrace a hope and look forward beyond your now.” - P18, Pastor, environmentalist}

\textit{“I feel I can hold two truths…and in Silicon Valley…the operating systems undergirding how we come to know reality and truth are so different [than my understanding]….truth in my religious tradition of Judaism has 49 faces. Truth is not singular. It’s not a fact. Truth is like this really big, mystical, embodied, intellectual, mind, body soul thing…” - P7, Researcher, Jewish }

\textit{"I think we have to be ok with not knowing and that is like…the opposite of most technology but I want technology to tell people, especially kids, but also adults, not everything has a tweetable  answer."  - P15, Tech worker, Pagan community leader}

Embracing wonder and uncertainty, for participants, also gave them hope for a future where technology’s current assumptions could be challenged and improved to center human flourishing. 

\section{Discussion}
From our interview study, we elicited 1) how religious people conceptualize secularity in technology and resulting unintended harms and 2) how religious epistemology shapes their technoreligious practices, boundaries with, and desires from technologies.  

But why should the HCI community care about this? And what can, or should, be done?

The core of Human–Computer Interaction (HCI) is improving how people engage with technology, designing systems that are intuitive and accessible for all users. Research shows that effective design requires understanding the mental models and metaphors people bring to technology \cite{norman1999affordance, norman2013design, vyas2006affordance, gibson2014ecological}. For many, religion is a central mental model for interpreting the world, making the study of religion and technology vital to advancing HCI’s core mission.

Yet religion remains understudied in computing. Past work documenting the lack of HCI and religion research suggests it may be underexplored due to its complexity and because engaging with it is seen as “complicated”, “difficult”, or “taboo” \cite{buie_spirituality_2013, wolf_still_2024}. But, we argue, this doesn’t mean it should be ignored. If anything, continuing to ignore religion could lead to more unintended harm which impact everyone -- religious or not. 

Technologies embody values stemming from the context of their development and their designers' positionalities —a notion deeply present in HCI research in frameworks like Winner's "Do Artifacts Have Politics" and Friedman's value-sensitive design, as well as empirical work towards technologists' formative positionalities \cite{winner2017artifacts, friedman_value-sensitive_1996, otuu_how_2025, scheuerman_products_2024}. If technology impacts all facets of life, and HCI researchers and technologists want human life to be better for it, the design, development, deployment, and evaluation of technology must contend with the holistic person – which includes religion for a majority of people \cite{fahmy_how_2025}. 

HCI has an exciting opportunity to deepen its understanding of the user by engaging with religion as an additional lens and site of innovation. In particular, we call for practitioners and researchers to see religion as an additional and valid sociocultural resource for consideration in technology design and to develop religious literacy to access this resource. 

But we know contending with religion is a large ask. We seek to make this task more tenable and to reduce the barriers of religion and HCI intersection work brought up in past literature \cite{buie_spirituality_2013, wolf_still_2024}. To do so, we distill our findings into meaningful and actionable design values for technologists (\ref{5.1}), and outline details and examples of what technologists in different focus areas can gain from religious literacy (\ref{5.2}). 

\subsection{Religiously informed and compatible design values}\label{5.1}
We propose six religiously informed design values, drawing on value-sensitive design \cite{friedman_value-sensitive_1996, ghoshal2023design}. These values are not prescriptive but serve as theoretical tools distilling lived religious experience into generative directions for technologists. We develop these values largely from our participants’ desires from technology (\ref{4.4}), contextualized by their current and past experiences with technology and situated in relevant literature and examples.

\subsubsection{Wonder: Looking beyond the known to the beauty of uncertainty}
Wonder, as an emotional experience, challenges technology design to move beyond certainty and the instant delivery of “accurate” information, instead inviting users to embrace emotion and uncertainty—qualities participants felt religion nurtures. Designing for wonder reframes technology not as a tool for definitive answers, but as a space for reflection, speculation, and openness to the unknown. Such systems might refuse to answer certain questions, encouraging users to seek people, sources, or their own interpretations, and to sit with the feelings that emerge. In this way, they cultivate emotionally aware sensemaking, a process well-studied in sociotechnical systems \cite{rumorssensemaking, emotionsensemaking}. Extending this logic beyond information retrieval, we might also ask how wonder could reshape immersive environments like VR: rather than striving for the most “realistic” replication of the physical world, what possibilities emerge when users are invited to inhabit new, unknowable worlds—digital spaces where not everything can be traversed, grasped, or fully known?

\subsubsection{Humility: Acknowledging limitations of technology}
Humility calls on technologists and their artifacts to foreground their limitations. We know from previous work that more data doesn’t always mean more insights or a better model \cite{thompson2021deep, boyd_six_2011}. We also know that several technologies posed as solutions, from the One Laptop Per Child \cite{ames_charisma_2019} to deployments of “AIs” in stores that were actually real-time human labelers \cite{Bitter2024Amazon}, did not live up to their hype \cite{bender2025ai}. Humility requires technologists to act more transparently. It also requires acceptance of the fact that “wicked problems” will have deeply multifaceted, non-singular solutions that are sociotechnical, not technosolutionist \cite{kuznetsov_study_2018, rittel_dilemmas_1973, wiltse_wicked_2015}. 

Although technology has much to offer, it cannot achieve what it aspires to without other ways of knowing and communities. Embracing humility to this reality opens space for community-driven codesign practices such as Taiwan’s g0v civic hacking \cite{chang2024temporal, lee2020nobody} and Japan’s SusHiTech movement \cite{SusHiTechTokyo2025}, which both combine community civic engagement, governmental collaboration/support and, as noted by Ecklund et al \cite{ecklund2019secularityFINAL} a culture where religion is a fuller participant in technical life. These projects harness horizontal connections between technologists and communities often forged partly through comembership of religious and other civil institutions to enact codesign values. Far from undermining technical progress, these practices seek to make technology sustainable through sociotechnical community integration.

\subsubsection{Space: For rest, reflection, and spiritual work}
In many religions, rest and reflection are held as sacred. But Silicon Valley and, likewise, academic research, are associated with a “grind mentality” and work ethic that has harmed workers across the supply chain \cite{gray2019ghost, McKendrick2019}. Much technology is based on a logic of productivity, even if it comes at the cost of unjust transitions of workers being automated or surveilled into unrelenting schedules \cite{levy2022data}. Other technologies often benefit from addictive behaviors and algorithms – taking away time from human connections \cite{Morell2025TechExit, costello2023algorithms, alter2018irresistible}. And while some solutions seek to disrupt tech addiction, these are not embedded values in technologies, but rather mitigations.

What might it look like to have technology experiences that encourage the preservation of human time? Things like noise canceling headphones and sensory friendly virtual experiences are a path towards such intentional spaces. But religious stakeholders have been making and preserving these spaces, from Sabbath lamps to Buddhist monks fine-tuning AI for their meditative practices, for millennia. HCI has an opportunity to learn and take inspiration from this. Many of these spaces are for spiritual, inner work, which our stakeholders are careful of having boundaries with technology. Such insights may inform and build solidarity between religious stakeholders and individuals performing creative and identity work \cite{lutz_were_2024, simpson2021you, lee2022rethinking, simpson2023rethinking} under infringement of technology. 

\subsubsection{Embodiedness: Transcendence through multisensory and proprioceptive experience}
Technology enables deeply human experiences, from technoreligious practices to broader instances of connection and community facilitation. These human experiences are deeply relational. But not all of them can be datafied and for many, technology is not a complete substitute. Religious perspective offers additional insight into some of the most powerful embodied experiences such as the transcendence experienced when memorizing and reciting the Quaran or embarking on a holy pilgrimage. If technologists embrace this notion that so many experiences cannot be fully digitized, it may open opportunities to other interaction modes which can serve as additive, extending experiences. As the HCI community already thinks about embodied interaction and embodiment \cite{dourish1999embodied, antle2011embodied, dourish_where_2001}, what might religious contexts add? For example, instead of mimicking weekly worship, are there experiences like exploring religious sites and ruins that are more rewarding and beneficial to develop in VR? What can multisensory experiences and ones involving natural (i.e. plant) interfaces  learn from experimenting with religious notions of embodiment? These case studies may enable new and exciting insights to creatively design for physical experiences and norms around co-presence in virtual worlds \cite{venkatraj_shareyourreality_2024, zojaji_join_2024}.

\subsubsection{Community: Preserving and enabling more togetherness}
Community is a central aspect of religious experience and is often seen as what distinguishes it from spirituality more broadly \cite{zinnbauer2015religion}. Religious participants emphasize how formal systems including technology and capitalism can undermine communal ties. To this end, we advocate for technologists to embrace prosocial agendas in technology design and evaluation \cite{harvey_hci_2014, weyl_prosocial_2025, lee_we_2023} – an agenda past HCI work has called to incorporate religion \cite{naqshbandi_making_2022}. Such prosocial agendas would focus on preserving and enhancing community relationships, seeing the strength of such relationships and the co-consumption of content, rather than individual attention, satisfaction or reward, as the target of design.

This also means considering prosocial experiences that extend community, such as experimenting with more experiences specifically designed for religious and spiritual co-presence of worship instead of meetings. Insights from these experiences, which are already nascent in many parts of the world \cite{the_economist_astrology_2025, Zaffar2025} and through religious community’s retrofitting of technologies from Zoom to Roblox \cite{therobloxianchristians2025, Kho2024RobloxFilipinoCatholics}, could inform other prosocial nonreligious community experiences. 

\subsubsection{Eternity: Escaping the cage of present}
Etymologically secularism is dedicated to an orientation of the present age, whilst religious orientations are invested in notions of eternity \cite{keane2000secularism}. This is another way secular embeddings manifest in technology, with short-term profits a well-documented priority over long-term sustainability of many technologies, both digital and physical \cite{arogyaswamy2020big, vrikki2024measuring, Zewe2025_GenAI_EnvironmentalImpact}. This concern is shared by the sustainable HCI community \cite{sustainableHCI1} and many religious stakeholders. 

Many interaction modalities focus on instant gratification over deeper, enduring experiences, a tension for religious stakeholders. Like the “slow technology movement" \cite{odom2012slow, postgrowth2025}, religious stakeholders advocate for technologies that encourage reflection and slowing down both use and consumption. HCI researchers have considered temporality in interaction, such as how many communities experience time (and therefore needs for technologies) differently \cite{norris_people_2022, wilner_its_2023, futureCHI}. Religious experiences, concerns, and negotiations around time add a promising new lens to this space. 

Embracing eternity suggestions deliberate practices of extending time horizons backwards and forwards, futuring grounded in tradition. Such practices are core in Asian and indigenous traditions, such as the recent application in Japan of the "Iroquois” 7 generations mechanism of role playing ancestors and descendants for dialog on planning \cite{saijo2020future}. The long sweep of history embedded in religious traditions and texts can thus be a resource for orientation towards the future many technologists call for but have struggled to embody \cite{macaskill2022we}.

Finally, designing for eternity means grappling with the full lifecycle of technological artifacts, from material extraction to eventual disposal. Religious traditions offer models for stewardship that extend beyond individual lifespans, suggesting technologies designed not just for current users but for the communities and ecosystems that will inherit them. This temporal expansion situates the present moment within larger patterns of care and responsibility that honor both past wisdom and future flourishing. 

\subsection{Future work towards religiously informed and literate computing}\label{5.2}
To engage with our design values, we advocate for technologists to develop religious literacy, the competency to understand how religion intersects with technology in political and social ways \cite{chan_religious_2021, hannam_religious_2020}. We hope this helps to facilitate closing the research gap at the intersection of HCI and religion, of which we provide examples of future needed work for communities of practice. 

\subsubsection{Fostering religious literacy in technologists}
Religious literacy is the competency to understand religion as a sociocultural resource, human phenomenon, and epistemological framework in the world \cite{chan_religious_2021, hannam_religious_2020}. We propose religious literacy as an additional sociocultural and theoretical framework in computing, aligning with other examples in HCI literature. Logistically, such literacy could be fostered through adding religion as a unit to engineering ethics courses \cite{horton2022embedding, harris1996engineering} or integrating it into workplace inclusion \cite{grim2016socio, grim2015modern}. But what does gaining religious literacy grant technologists?

Religious literacy isn’t just about designing technologies to be more inclusive of religious use cases. It is a commitment to valuing religion as an additional and valid sociocultural resource. Pragmatically, this may be useful when deploying technologies in majority religious parts of the world or engaging religious communities as user populations in research and testing. 

But more substantially, religious literacy grants new theoretical tools to examine technology. In contending with issues of tech ethics, Ames and colleagues applied religious metaphors to engineering ideological commitments (and shortcomings) \cite{ames_worship_2015}. Meanwhile, in Tech Agnostic, Epstein leverages theology to document the arrangements of power and rituals within technology – and how to question them \cite{epstein2024tech}. What might it mean for more scholars to engage with religiously informed theoretical work or empirical contexts? Are there benefits to framing the “AI race” through the lens of idolatry? To understanding systems through a lens of polytheism? Religious literacy is about understanding religion as a cosmology of beliefs, people, organizations – and power. Through doing so, technologists and researchers can see that technology is shaped by similar forces and that both religion and technology influence sociotechnical systems and human life. Understanding this can offer salient new insights and actionable directions, as our empirical work has provided via religiously informed design values and future research directions. 

\subsubsection{Key communities of practice for future work}
Based on our work, we highlight key communities of practice within HCI relevant towards an initial research agenda of religiously informed and literate computing. In particular, we believe these communities and areas could benefit from exploring our design values and from testing and researching technologies in religious contexts.

\begin{itemize}
    \item \textbf{Education:} Much of HCI contends with education such as interactive educational technologies, digital literacies, and computing education for children and adults of all ages. Religion is one of the oldest and most established institutions of education in the world, particularly of continuous adult education \cite{saul1997adult}. Are there important empirical insights and design values that could be inferred from religious learning to benefit more broad educational technologies? What different affordances do religious educational institutions need? And, in educational efforts such as AI literacy for children to senior citizens, could houses of worship serve as an additional vector for reaching more people?
   
    \item  \textbf{Digital safety and FATE research:} A variety of human identity biases in sociotechnical systems have been extensively examined by researchers, often with the hopes of reducing allocational and representational harms in sociotechnical systems \cite{ghosh_generative_2024, ghosh_i_2024, dev_measures_2022, frameworkscheu2021} – often described in the “FAccT” or “FATE” research umbrellas. We hope to see religion as an additional lens in this work. Additionally, as digital safety work like red-teaming engages with outside experts (i.e. psychologists) and contends with sociocultural expertise \cite{ahmad2025openai, gillespie2024ai, zhang2024human}, we hope to see faith leaders as additional experts in such activities and as potentially untapped resourcing for digital safety work.
   
    \item  \textbf{Physical, ubiquitous, and immersive computing:} Houses of faith can serve as new testing grounds for immersive and multisensory computing experiences towards religious expression. Although some religious groups are experimenting with immersive experiences, such technologies have not been extensively researched for and within these contexts. What can be learned from designing such experiences in religious contexts? 
    
    \item  \textbf{Computer support collaborative work (and worship) (CSCW)}: CSCW, a subfield of HCI, is dedicated to how different notions of work aremediated by collaboration and technology. Religious communities are understudied as is the emergent notion of “spiritual work” \cite{welwood_principles_1984, wyche_extraordinary_2009} and how religious people conceptualize types of and boundaries with work, such as rest. Research in religiously contextual work may add imperative insights to navigate disruptions from our current AI era of automation in healthy, holistic ways. In considering notions of computer-supported collaborative worship, what other holistic activities may be enabled?
    
    \item   \textbf{AI human interaction:} As human-AI interaction becomes a growing focus, the notion of individual companions and agents may be worth challenging. Are there other paradigms of interaction that may work for more mental models of humans? Additionally, when considering AI’s mediations in personal practices, including religious ones, what interactivity modes may emerge? 
    
    \item \textbf{Policy and regulation:} Politically-organized religious citizens have long been studied as the central influence on many areas of technology policy and standards, especially related to reproduction and content standards in broadcast media \cite{hoover1997history, pavolini2017mapping, beckford2007religion}. But much of this research has not made it to HCI, with neither past systematic review \cite{buie_spirituality_2013, wolf_still_2024} covering this intersection of religion, politics, and technology. This is a pertinent space for future work, especially given recent religious political groups and religious organizations like the Vatican taking interest in AI regulation and policy \cite{zengarini_world_2025, Vatican2025AntiquaEtNova}.
      
\end{itemize}

This is not an exhaustive list of stakeholder groups we think should explicitly contend with religion. Rather, we hope these concrete examples and our design values (\ref{5.1}) may empower and inspire researchers reckoning with sociotechnical systems to integrate religious people and considerations into their work as an additional humanistic lens. 

\subsubsection{Overcoming barriers to facilitate and increase religious research in HCI}
It is important, however, to address an elephant in the room: the long-standing tensions and sensitivity, noted by previous research surveys \cite{buie_spirituality_2013, wolf_still_2024} around research in religion and spirituality in HCI. While we have discussed extensively why overcoming this sensitivity is important through positioning religion as an additional sociocultural resource, it is also important to highlight that it is feasible. The assumption of inevitable conflict and tension is largely driven by an Anglo-centric frame; in other parts of the world, both our research and previous studies have found the relationship between these fields is much more complementary \cite{ecklund2019secularityFINAL, fahmy_23_2025}. Furthermore, as attention from religious leaders towards technology issues grow \cite{Vatican2025AntiquaEtNova, zengarini_world_2025} and as policy and technology leaders increasingly look to religious leaders for guidance in this space \cite{YohnTaylor2025}, we expect funding in this area to rise and the potential impact of such research to escalate. Thus, we hope that many of the previously-highlighted impediments to work at this intersection are easing and we can see a flowering of the critically-needed contributions like those we highlighted.

We hope that our work has shown how researchers can move past some of these barriers by providing an empirically grounded design values and insights that show how technologists can leverage religion as an additional and valid sociocultural resource when designing technologies for the holistic user. 

\section{Limitations}\label{limitations}
As with all qualitative work, there are limitations to the claims we can make. Firstly, our sample is not representative of all religious people, even majority religious. Along with selection bias, our interview pool lacks sufficient representation from many faiths, particularly Hinduism -- a large  limitation given the large size of this religion and its intersections with technology. Our pool also leans towards Abrahamic faiths, particularly Christianity (partially due to our US context). Additionally, many of our interviewees were researchers themselves and individuals who occupy roles in academia, think-tanks, and other types of knowledge work who had thought extensively about the intersection of technology and religion, adding further bias. Our recruitment too of many of these connections being from the interviewer's managers also colored results.

Religion is a deeply personal and intimate topic, but also radically stoked in culture, history, and geopolitics. This study and its insights are US-centric, due to the research team’s positionality and employment. Although a minority of stakeholders were based outside of the US, all interviews were conducted in English. This overall context and linguistic limitation does color our findings and their applicability in the international stage. 

Our work acts as a landscape study to document perceptions and desires of technology by religious stakeholders – hoping to inspire future work that incorporates religious literacy and expertise. We particularly hope future work will include stakeholders and perspectives we were less able to, particularly with more Eastern and polytheistic faiths.

\section{Conclusion}
We have provided an empirical landscape study of religious stakeholders’ perceptions of and desires from technologies and the technologists who design them. 

In doing so, we have surfaced key findings about how religious stakeholders perceive secular embeddings and assumptions in technology and how this has led to exclusion, particularly in larger “big data” waves of technologies. These exclusions have led to unintended harms to religious communities, as well as others and more importantly to crucial missed opportunities for technologists to serve the needs of religious communities and draw on their wisdom to shape technologies for all stakeholders. Religious stakeholders have deeply contextual relationships with technology in not only technoreligious practice but in daily life – these boundaries and negotiations further shape their desires from technology and the technologists making them. 

We have provided, from these findings, actionable design values to distill religious wisdom and context into technology. We hope these values and our call for increased religious literacy and areas of a research agenda within HCI will help to increase work towards making technology for the holistic user.

\bibliographystyle{ACM-Reference-Format}
\bibliography{MSRReligion, 2_religion}

\appendix

\end{document}